\documentclass[aps,prl,twocolumn,amsmath,amssymb,nofootinbib,superscriptaddress,floatfix,reprint,longbibliography]{revtex4-1}

\usepackage{amsmath}
\usepackage{amssymb}
\usepackage{amsthm}
\usepackage[pdftex]{color} %Titus
\usepackage{graphicx}
\usepackage{dcolumn} % Align table columns on decimal point
\usepackage{bm} % bold math
\usepackage{url}
\usepackage[colorlinks=true, urlcolor=blue, linkcolor=blue, citecolor=blue, pdftex]{hyperref}
\usepackage{longtable}
\usepackage{ulem} % to strike things out
\begin{document}

\title{Nature of unconventional pairing in the kagome superconductors AV$_3$Sb$_5$}

\author{Xianxin Wu}
\email{xianxin.wu@fkf.mpg.de}
\affiliation{Max-Planck-Institut f\"ur Festk\"orperforschung, Heisenbergstrasse 1, D-70569 Stuttgart, Germany}
\affiliation{CAS Key Laboratory of Theoretical Physics, Institute of Theoretical Physics, Chinese Academy of Sciences, Beijing 100190, China}

\author{Tilman Schwemmer} \affiliation{Institute for Theoretical Physics,
  University of W\"{u}rzburg, Am Hubland, D-97074 W\"{u}rzburg, Germany}

\author{Tobias M\"uller} \affiliation{Institute for Theoretical Physics,
  University of W\"{u}rzburg, Am Hubland, D-97074 W\"{u}rzburg, Germany}

\author{Armando Consiglio} \affiliation{Institute for Theoretical Physics,
  University of W\"{u}rzburg, Am Hubland, D-97074 W\"{u}rzburg,
  Germany}
\author{Giorgio Sangiovanni} \affiliation{
Institut f\"ur Theoretische Physik und Astrophysik and W\"urzburg-Dresden Cluster of Excellence ct.qmat, Universit\"at W\"urzburg, 97074 W\"urzburg, Germany}

\author{Domenico Di Sante} \affiliation{Department of Physics and Astronomy, Alma Mater Studiorum, University of Bologna, 40127 Bologna, Italy}
\affiliation{Center for Computational Quantum Physics, Flatiron Institute, 162 5th Avenue, New York, New York 10010, USA}

\author{Yasir Iqbal} \affiliation{Department of Physics and Quantum Centers in Diamond and Emerging Materials (QuCenDiEM) group, Indian Institute of Technology Madras, Chennai 600036, India}

  \author{Werner Hanke} \affiliation{Institute for Theoretical Physics,
  University of W\"{u}rzburg, Am Hubland, D-97074 W\"{u}rzburg, Germany}

\author{Andreas P. Schnyder}
\affiliation{Max-Planck-Institut f\"ur Festk\"orperforschung, Heisenbergstrasse 1, D-70569 Stuttgart, Germany}

\author{M. Michael Denner} \affiliation{Department of Physics, University of Zurich, Winterthurerstrasse 190, 8057 Zurich, Switzerland}

\author{Mark H. Fischer} \affiliation{Department of Physics, University of Zurich, Winterthurerstrasse 190, 8057 Zurich, Switzerland}

\author{Titus Neupert} \affiliation{Department of Physics, University of Zurich, Winterthurerstrasse 190, 8057 Zurich, Switzerland}

  \author{Ronny Thomale}
  \email{rthomale@physik.uni-wuerzburg.de}
\affiliation{Institute for Theoretical Physics,
  University of W\"{u}rzburg, Am Hubland, D-97074 W\"{u}rzburg, Germany}
\affiliation{Department of Physics and Quantum Centers in Diamond and Emerging Materials (QuCenDiEM) group, Indian Institute of Technology Madras, Chennai 600036, India}

\date{\today}

\begin{abstract}
The recent discovery of
AV$_3$Sb$_5$ (A=K,Rb,Cs) has uncovered an intriguing arena for exotic Fermi surface
instabilities in a kagome metal. Among them, superconductivity is
found in the vicinity of multiple van
Hove singularities, exhibiting indications of unconventional
pairing. We show that the sublattice interference
mechanism is central to understanding the formation of
superconductivity in a kagome metal. Starting from an appropriately
chosen minimal tight-binding
model with multiple van Hove singularities close to the Fermi level for AV$_3$Sb$_5$, we provide a random phase approximation analysis of
superconducting instabilities. Non-local Coulomb repulsion, the sublattice
profile of the van Hove bands, and the interaction strength
turn out to be the crucial parameters to determine the preferred
pairing symmetry. Implications for potentially topological surface
states are discussed, along with a proposal for additional measurements to pin
down the nature of superconductivity in AV$_3$Sb$_5$.
\end{abstract}

\maketitle

{\it Introduction.} The kagome lattice has become a paradigmatic
setting for exotic quantum phenomena of electronic
matter. This particularly applies to quantum magnetism, where the
large geometric spin frustration inherent to the corner-sharing
triangles promotes the emergence of extraordinary quantum
phases~\cite{RevModPhys.88.041002}. From an itinerant limit, electronic kagome bands are likewise
particular, as
they feature a flat band, Dirac cones, and van Hove singularities at
different fillings. The kagome flat band suggests itself as a natural
host for the
realization of
ferromagnetism~\cite{Mielke_1991,PhysRevLett.100.136404} or possibly
non-trivial topology~\cite{PARAMESWARAN2013816,PhysRevB.93.155155,YinJX2020,PhysRevLett.125.266403},
while the kagome Dirac cones have been proposed to be a promising
way to accomplish strongly correlated Dirac fermions~\cite{mazin} and turbulent
hydrodynamic electronic flow~\cite{disante}. The kagome lattice at van Hove filling
has been shown to be preeminently suited for the emergence of exotic
Fermi surface instabilities~\cite{PhysRevB.85.144402,PhysRevB.86.121105,PhysRevB.87.115135,PhysRevLett.110.126405}. Among others, this involves charge and spin density-wave orders with
finite relative angular momentum~\cite{PhysRevB.62.4880}. Moreover,
the kagome Hubbard model was first predicted to yield degenerate nematic instabilities
which can break point-group and time-reversal symmetry simultaneously~\cite{PhysRevLett.110.126405},
which has currently regained attention in the context of twisted bilayer graphene~\cite{Fernandeseaba8834}.

The recent discovery of AV$_3$Sb$_5$~\cite{PhysRevMaterials.3.094407} provides an instance of
kagome metals tuned to the vicinity of multiple van Hove
singularities. What further makes
them unique is the combination of metallicity, strong
two-dimensional electronic character, and significant electronic
correlations derived from the $d$-orbital structure of the Vanadium
kagome net.  KV$_3$Sb$_5$ was discovered to be a kagome superconductor
with $T_c=0.93$~K~\cite{PhysRevMaterials.5.034801}, along with RbV$_3$Sb$_5$ ($T_c=0.92$~K)~\cite{QiangweiYin.37403} and
CsV$_3$Sb$_5$ ($T_c=2.5$~K)~\cite{PhysRevLett.125.247002,zhao2021cascade}, where the latter was shown to rise up to
$T_c=8$~K under 2~GPa hydrostatic pressure~\cite{chen2021double,Zhang2021,Chen2021}. While the wheel of
experimental exploration is still in spin, certain tendencies about
the superconducting phase are starting to crystallize. The observed charge density wave (CDW) order \cite{jiang2020discovery}, interpreted
as a potential parent state for unconventional superconducting order~\cite{PhysRevLett.110.126405,Feng2021,Denner2021,Lin2021}, exhibits
indications for an electronically driven formation~\cite{Li2021}.
Specific-heat measurements suggest at least a strongly anisotropic
gap~\cite{PhysRevMaterials.5.034801}.
While a significant residual term from thermal conductivity suggests a nodal
gap~\cite{zhao2021nodal}, penetration depth measurements claim a
nodeless gap~\cite{Duan2021}. The dome shape suggests
unconventional superconductivity along with a large value of
$2\Delta/k_{\text{B}}T_{
\rm c}$, hinting at a strong-coupling
superconductor~\cite{Chen2021a}.

In this Letter, we formulate a theory of unconventional
superconductivity in AV$_3$Sb$_5$. In a first step,  we develop an
effective tight-binding model suitable for the analysis of pairing instabilities. In order to retain the necessary complexity of multiple van Hove singularities in the vicinity of the Fermi
level in AV$_3$Sb$_5$, we distill a six-band minimal model. In a second
step, we specify the interaction Hamiltonian. Due to matrix elements
implied by the sublattice interference mechanism~\cite{PhysRevB.86.121105}, which we review below, it is essential to
take non-local Coulomb repulsion into consideration. Over a large range of coupling strengths, we find dominant $f$-wave triplet superconducting
order, succeeded by $d$-wave singlet pairing for stronger coupling.
Throughout the phase diagram, the $p$-wave order stays subdominant but competitive.
Aside from this general trend, the detailed competition
between the different orders is crucially influenced by the location
of the Fermi level with respect to the multiple van Hove singularities and the nearest-neighbor (NN) Coulomb repulsion.

\begin{figure}[tb]
\centerline{\includegraphics[angle=270,width=1.0\columnwidth]{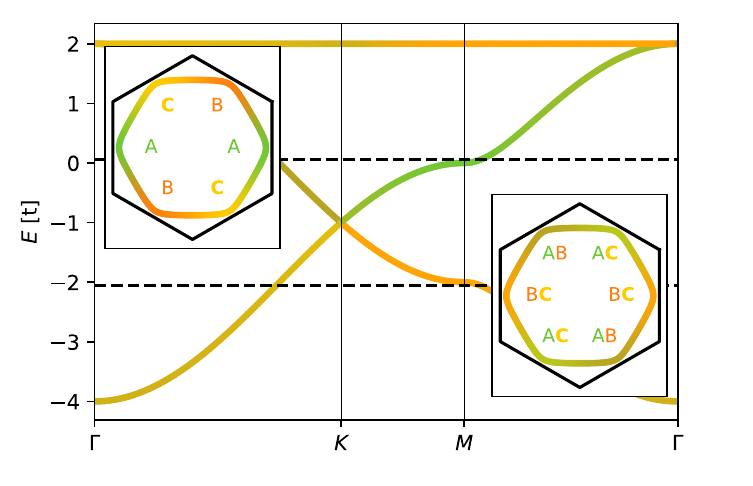}}
\caption{Generic three-band kagome dispersion exhibits two van-Hove singularities
  with distinct sublattice (A,B,C) decoration. Whereas the upper van Hove filling at $E
  = 0$ has a pure sublattice makeup (p-type, left inset), the lower van Hove filling
  at $E=-2\,t$ mixes different sublattice states (m-type, right inset). Energy is measured in units of the NN hybridization $t$.
 \label{fig:kagomesub} }
\end{figure}

{\it Sublattice decoration of kagome van Hove points.}
As opposed to related hexagonal van Hove singularities such as for the
bipartite honeycomb lattice, the kagome bands can host two different types
of van Hove singularities which we label as sublattice mixing (m-type) and
sublattice pure (p-type), characterized by odd and even parity at the M point, respectively.
This is illustrated in Fig.~\ref{fig:kagomesub} for the minimal kagome tight-binding model
with three distinct sublattice sites located on the 3f Wyckoff positions of the $P6/mmm$ space group.
The upper van Hove singularity ($E=0$) is of p-type, since the Fermi level eigenstates
in the vicinity of the three $M$ points are localized on
mutually different sublattices (left inset). By contrast, the lower
van Hove filling ($E=-2\,t$) has mixed sublattice character and thus is of m-type, with the
eigenstates equally distributed over mutually different sets of two
sublattices for each $M$ point (right inset).
These distinct sublattice decorations have a strong impact on the
nesting properties (see Sec. II of supplementary materials (SM)~\cite{SM}). Since p-type van Hove points do not
couple to each other via local interactions, the inclusion of at least
NN Coulomb repulsion is quintessential to adequately
model interacting kagome metals close to p-type van Hove filling~\cite{PhysRevB.86.121105,Zhao2021}.

{\it Effective model.}
The ab-initio band structure of AV$_3$Sb$_5$ matches well with ARPES measurements below the CDW transition temperature, even though the corresponding density functional theory (DFT) calculations are performed neglecting the star-of-David-type structural distortion~\cite{PhysRevLett.125.247002,tan2021charge}.
Due to the multiple sublattices and the large number of contributing orbitals from both V and Sb in the vicinity of the Fermi level, a reduction to an effective model is a prerequisite to any analysis of many-body instabilities.

\begin{figure*}[tb]
\centerline{\includegraphics[angle=270,width=1.0\linewidth]{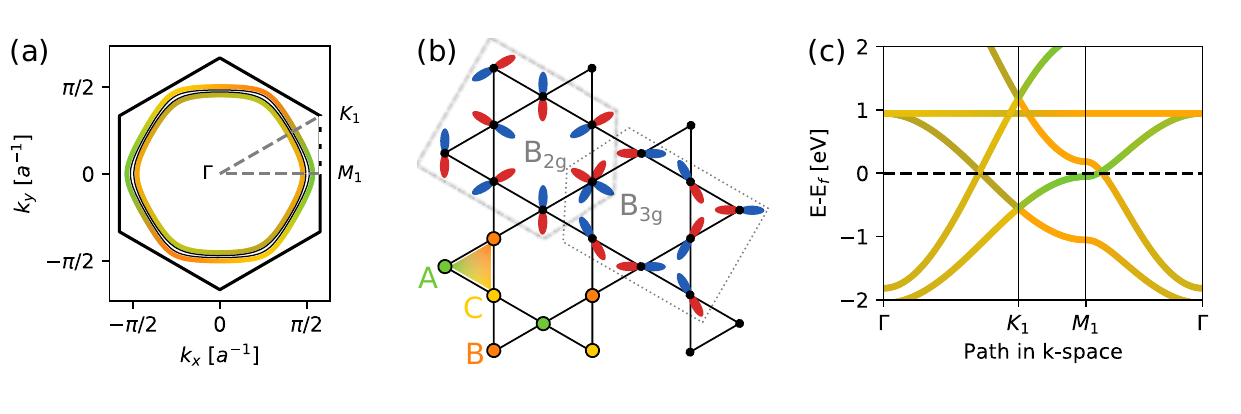}}
\caption{(a) Fermi surfaces of the minimal two-orbital tight-binding
  model [Eq.(\ref{eq:H-dxzdyz})] for a Vanadium
kagome net. Both a p-type and m-type van Hove filling are present
nearby the Fermi level, indicated by their respective orbital
color. (b) Real-space structure of the kagome V planes. The sign
structure (blue/red) and spatial orientation of the $B_{2g}$ and
$B_{3g}$ orbitals is sketched on different lattice sites. The red,
yellow and orange coloring indicates the three kagome sublattices.
(c) Band structure along the high-symmetry path depicted by the dashed
line in (a). The band coloring indicates the sublattice support of the
momentum eigenstates.
%\sout{Parameters for these plots are\(t_{\mathrm{xz}}=1\,\text{eV}\), \(t_{\mathrm{yz}}=0.5\,\text{eV}\),
%\(t' = 0.002\,\text{eV}\), \(\epsilon_\mathrm{xz} = 2.182\,\text{eV}\), and
%\(\epsilon_\mathrm{yz} = -0.055\,\text{eV}\).}
 \label{fig:TB_model} }
\end{figure*}

The layered structure of AV$_3$Sb$_5$, together with the large transport anisotropy
of \(\rho_c/\rho_{ab}\approx600\)~\cite{PhysRevLett.125.247002} (for
CsV$_3$Sb$_5$) allows us to constrain
ourselves to the two-dimensional V-Sb kagome plane. Analyzing the Fermi level at \(k_z = 0\) by means of density
functional theory, we find three distinct Fermi surfaces in AV$_3$Sb$_5$:
(i) a pocket composed of Vanadium $d_{\mathrm{xy}}, d_{\mathrm{x}^2-\mathrm{y}^2}, d_{\mathrm{z}^2}$ orbitals in proximity to a
p-type van Hove singularity, (ii) two additional pockets composed of Vanadium $d_{\mathrm{xz}}$,$d_{\mathrm{yz}}$
orbitals in proximity to another p-type and m-type van Hove singularity above
and below the Fermi level, respectively (Fig.~\ref{fig:TB_model}), and (iii) a circular
pocket around  \(\Gamma\) formed by Antimony $p_{\mathrm{z}}$-orbitals. Note that
(i) and (ii) do not hybridize due to opposite $M_z$ eigenvalues and
the symmetry-wise allowed hybridization of (ii) and (iii) is
parametrically weak. These features are not particularly sensitive to
spin-orbit coupling, which is hence not further considered in the following.

For the effective model, we restrict ourselves to the Fermi pockets
(ii) for three reasons. First, the pockets in (ii) carry the dominant
density of states at the Fermi level. Second, we preserve the complexity of
multiple van Hove singularities of p-type and m-type in our minimal
model. Third, upon comparison to the ab-initio band structure, our minimal model manages to correctly
capture all irreducible band representations at the high symmetry
points in the Brillouin zone. The constituting $d_{\mathrm{xz/yz}}$ orbitals belong to the $B_{2g/3g}$ irreducible representations of the site symmetry group $D_{2h}$ for the 3f Wyckoff positions [Fig.~\ref{fig:TB_model}(b)], forming a set of bands with opposite mirror eigenvalues along the $\Gamma$--$M$ line.
These bands give rise to a mirror-symmetry-protected Dirac cone on the
$\Gamma$--$M$ line and hence, an upper and lower van-Hove filling with
opposite sublattice parity (Fig.~\ref{fig:TB_model}). Employing the $D_{6h}$ point group symmetry, our corresponding
effective six-band Hamiltonian can then be derived as
\begin{eqnarray}
%\begin{split}
    H &=& \
    \sum_{\mathbf{k}i\alpha} \epsilon_{\alpha}^{} \
    c_{\mathbf{k}i\alpha}^{\dagger}c_{\mathbf{k}i\alpha}^{} 
    -\sum_{\mathbf{k}ij\alpha} t_{\alpha}^{}
    \Phi_{ij}^{}(\mathbf{k})^{}  \
    c_{\mathbf{k}j\alpha}^{\dagger}c_{\mathbf{k}i\alpha}^{} \nonumber\\
    &&- t' \sum_{\mathbf{k},ij} \
    \Phi_{ij}^{}(\mathbf{k}) s_{ij}^{} \
    (c_{\mathbf{k}j\mathrm{xz}}^{\dagger}c_{\mathbf{k}i\mathrm{yz}}-c_{\mathbf{k}j\mathrm{yz}}^{\dagger}c_{\mathbf{k}i\mathrm{xz}}) ,
%\end{split}
\label{eq:H-dxzdyz}
\end{eqnarray}
where
($i = \mathrm{A},\mathrm{B},\mathrm{C}$ and $\alpha = \mathrm{xz},\mathrm{yz}$). The crystal field splitting is denoted by $\epsilon_{\alpha}$,  where the operator $c_{\mathbf{k}i\alpha}^{\dagger}$ ($c_{\mathbf{k}i\alpha}^{}$) creates (annihilates) an electron with momentum $\mathbf{k}$ of sublattice $i$ in orbital $\alpha$.  The lattice
structure factors read
$\Phi_{\mathrm{AB}}(\mathbf{k}) = 1 + e^{-2i \mathbf{k}\cdot \mathbf{a}_1}$,
$\Phi_{\mathrm{BC}}(\mathbf{k}) = 1 + e^{-2i \mathbf{k}\cdot \mathbf{a}_3}$, and
$\Phi_{\mathrm{AC}}(\mathbf{k}) =  1+e^{-2i \mathbf{k}\cdot \mathbf{a}_2}$
obeying the hermiticity condition
$\Phi_{ij}^{}(\mathbf{k})=(1-\delta_{ij})\Phi_{ji}^*(\mathbf{k})$,
where the sublattice-connecting vectors are denoted by
$\mathbf{a}_{1,2} = \left(\sqrt{3}/2, \pm 1/2\right)^{\mathrm{T}}$ and
$\mathbf{a}_{3} = \left(0,-1\right)^{\mathrm{T}}$. The second term represents
the intra-orbital NN hoppings on the kagome lattice with
two distinct amplitudes $t_{\alpha}$, while the third term describes
NN inter-orbital hopping amplitude $t'$. The non-trivial
transformation properties of the $d_{\mathrm{xz}}$ and
$d_{\mathrm{yz}}$-orbitals under the site-symmetry group result in a
non-trivial sign structure for the third term, described by $s_{\mathrm{AC}}=s_{\mathrm{CB}}=-s_{\mathrm{AB}}$ and $s_{ij}=-s_{ji}$.
We approximately fit our model to the ab-initio band structure
(see Sec. IV of SM) and obtain the parameters
\(t_{\mathrm{xz}}=1\,\text{eV}\), \(t_{\mathrm{yz}}=0.5\,\text{eV}\),
\(t' = 0.002\,\text{eV}\), \(\epsilon_\mathrm{xz} = 2.182\,\text{eV}\), and
\(\epsilon_\mathrm{yz} = -0.055\,\text{eV}\). The corresponding band structure and Fermi surfaces are shown in Fig. \ref{fig:TB_model}.

{\it RPA analysis.}
For the electronic interactions, we consider multi-orbital
density-density type interactions up to NNs
\begin{equation}
    \begin{split}
H_{\text{int}}=&U\sum_{li\alpha}n_{li\alpha\uparrow}n_{li\alpha\downarrow} + U'\sum_{li,\alpha<\beta}n_{li\alpha}n_{li\beta} \\
 &+J\sum_{li,\alpha<\beta,\sigma\sigma'}c^{\dag}_{li\alpha\sigma}c^{\dag}_{li\beta\sigma'}c_{li\alpha\sigma'}c_{li\beta\sigma}\\
&+J'\sum_{li,\alpha\neq\beta}c^{\dag}_{li\alpha\uparrow}c^{\dag}_{li\alpha\downarrow}c_{li\beta\downarrow}c_{li\beta\uparrow}+\sum_{\langle ll'\rangle ij\alpha\beta}V_{\alpha\beta}n_{li\alpha}n_{l'j\beta},
\label{interaction6band}
\end{split}
\end{equation}
where $n_{li\alpha}=n_{li\alpha\uparrow}+n_{li\alpha\downarrow}$ and $l,l'$ is an index for the unit cell. $U$, $U'$, $J$, and $J'$ denote the onsite
intra-orbital, inter-orbital repulsion, Hund's coupling, and
pair-hopping terms, respectively. $V_{\alpha\beta}$ denotes the
repulsion between NN sites. In the following we adopt
the parameterization $U = U'+2J$, $J = J'$ with $J = 0.1\,U$ and
$V_{\alpha\beta}=0.3\,U$ $\forall \alpha,\beta$  consistent with our
ab-initio cRPA estimates for a target manifold comprising V-3$d$ and
Sb-5$p$ orbitals~\cite{ferdiPRB}. An extensive ab initio study of
interactions and their dependence on the effective low-energy
model will be presented elsewhere.

\begin{figure}[t]
\centerline{\includegraphics[angle=270,width=1.0\columnwidth]{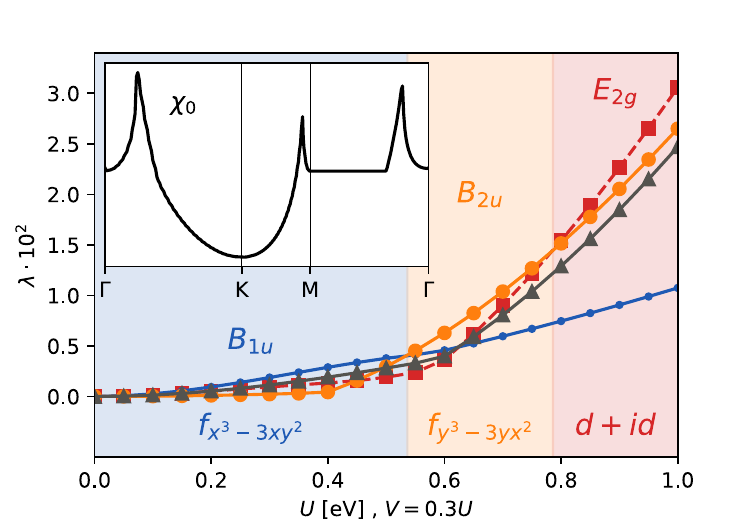}}
\caption{
Pairing-strength eigenvalues $\lambda$ for the dominant instabilities
as a function of $U$ ($V=0.3U$, $J=0.1U$) in Eq. (\ref{interaction6band}).
Continuous (dashed) lines indicate triplet (singlet) pairing. Two
distinct $f$-wave solutions, $B_{1u}$ (blue dots) and $B_{2u}$
(large orange dots), dominate for smaller interaction scales. The
$d$-wave $E_{2g}$ solution dominates for larger $U$. The $p$-wave
$E_{2u}$ solution (gray triangles) is subleading, but competitive at
all $U$. The upper left inset depicts the largest eigenvalue
trajectory of the bare susceptibility
$\chi_{0}(\bm{q})$ along the high-symmetry path indicated in
Fig.~\ref{fig:TB_model}(a).
 \label{rpapairingV03} }
\end{figure}

The inset of Fig.~\ref{rpapairingV03} displays the leading eigenvalue
of the bare susceptibility $\chi_{0}(\bm{q})$ along high-symmetry
lines. It is mainly attributed to the $d_{\mathrm{yz}}$ orbital and features
three prominent peaks. The largest two are located proximate to the
$\Gamma$ point, while the peak close to $M$ is suppressed through
sublattice interference. Including onsite and NN interactions at the
RPA level, these peaks get significantly enhanced in the spin as well
as charge channel. Note that, indeed, we find the charge susceptibility
at the verge of diverging around the $M$ point for strong NN
repulsion, hinting at an incident CDW instability.

Below the critical interaction, superconductivity emerges, triggered by
charge and spin fluctuations~\cite{Scalapino1986,Onari2004,Graser2009}. The obtained pairing eigenvalues as a
function of $U$ are displayed in Fig.~\ref{rpapairingV03}. For
$U<0.54$ eV, pairing on the p-type Fermi sheet from the p-type van Hove band with $B_{1u}$ ($f_{x^3-3xy^2}$-wave) symmetry is favored and $E_{1u}$ ($p$-wave) and $E_{2g}$ ($d$-wave) pairings are subdominant.
Increasing the coupling results in a rapid increase of the $B_{2u}$ ($f_{y^3-3yx^2}$-wave) and $E_{2g}$ pairings on the Fermi sheet from the m-type van Hove band,
where the spin-triplet solution still dominates slightly.
Upon further increase of the interaction strength, the $d$-wave pairing on this Fermi sheet becomes dominant.
Meanwhile, the $E_{1u}$ pairing is subdominant. Varying the ratio of $V/U$ ($0.2<V/U<0.35$) does not qualitatively change the above results in the six-band tight-binding model (see Sec. IV of SM). Note that we have also performed calculations with a seven-band tight-binding model including the circular pocket around $\Gamma$ \cite{Sakakibara2020} and find that it has negligible effect on the pairing (see Sec. IV of SM).

\begin{figure}[tb]
\centerline{\includegraphics[angle=0,width=1.0\columnwidth]{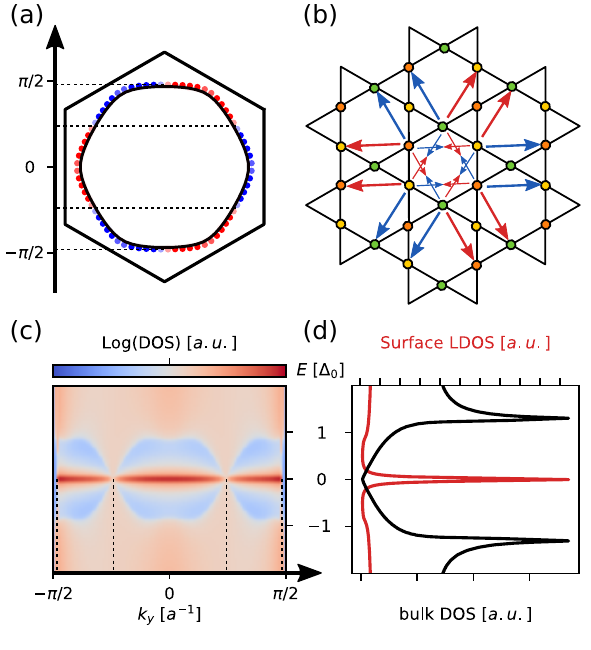}}
\caption{The inter-sublattice triplet $f_{x^3-3xy^2}$-wave pairing function in (a) momentum and (b) real space, where the superconducting order parameter changes sign under 60$^{\circ}$ rotation.
(c) Edge spectra for open boundary conditions along $x$ direction; zero-energy Andreev bound states appear between the projections of nodal points in the $f$-wave pairing. (d) Density of states for the bulk $f$-wave pairing (black) and local density of states at edges from Andreev bound states (red).
 \label{fwavepairing} }
\end{figure}

 We further analyze the harmonic fingerprint of the obtained pairings. The $f_{x^3-3xy^2}$-wave pairing is dominated by the sublattice-pure $d_{\mathrm{yz}}$ Fermi surface and the corresponding gap function in $k$-space is shown in Fig.~\ref{fwavepairing}(a), where there are line nodes along $\Gamma$--$K$ and the superconducting gap changes sign under 60$^{\circ}$ rotation.
 The corresponding real-space pairing is displayed in
 Fig.~\ref{fwavepairing}(b), which represents a spin-triplet
 sublattice-triplet pairing between $d_{\mathrm{yz}}$ orbitals on the next-nearest-neighbor (NNN) sites. This pairing is promoted by the
 effective interaction between the NNN sites from the second-order
 effect of NN repulsion, an effect which is robust to including diagrams beyond the RPA approximation as one would do in a functional renormalization group (fRG) study~\cite{Honerkamp2008,Wu2020PRB}. %Maybe we add some calculation to the SI here
 Counter-intuitively, we find that both $B_{2u}$ and $E_{2g}$ states, favored at larger $V$, are dominated by
 the sublattice-mixing $d_{\mathrm{xz}}$ Fermi surface and attributed to the pairing between
 $d_{\mathrm{xz}}$ orbitals on the NN sites. The $B_{2u}$ pairing, different from the $B_{1u}$ pairing, possesses line nodes
 along the $\Gamma $--$M$ direction. We expect the two-fold degenerate $E_{2g}$ pairing instability to form a $d+id$ state below $T_c$ in order to
 maximize condensation energy, which spontaneously breaks
 time-reversal symmetry. Notably, as a particular feature of the
 kagome lattice, the pairings between NN sites are promoted by the inter-orbital NN
 repulsion: although there is a direct repulsion between the NN sublattices, the second-order contribution via the other
 sublattices can be attractive. Once it overcomes the direct repulsion
 term, the effective interaction between NN sites
 becomes attractive and can promote NN pairing. Furthermore, note that the superconducting gap in the obtained states
for our minimal model are either dominant on the $d_{\mathrm{yz}}$ or $d_{\mathrm{xz}}$
Fermi surface, which can be attributed to the assumed weak inter-orbital
hopping.

{\it Topological properties of the pairing states.}
Our minimal-model analysis is dominated by an $f$-wave state for weak
coupling. Combined with the observation that the band renormalization
in ARPES appears moderate, $f$-wave order could be a favored
candidate for the nature of pairing in AV$_3$Sb$_5$. For time-reversal-invariant superconductors, the topological criterion
about zero-energy Andreev bound states on edges is determined by
winding numbers~\cite{Sato2011,Schnyder2012}. For both $f$-wave
pairing states emerging in our analysis, each node carries a winding number of $+1$ or $-1$. If we impose open boundary conditions, where the projections of nodes with opposite winding number do not overlap, a zero-energy flat band connecting the projections of nodes is created.
For illustration, we present the surface spectrum of the
$f_{x^3-3xy^2}$-wave state with open boundary conditions along the $x$
direction in Fig.~\ref{fwavepairing}(c). The corresponding local
density of states features a sharp zero-bias peak, shown in
Fig.~\ref{fwavepairing}(d) which could be observed at corresponding
step edges in STM measurements.
A similar analysis can likewise be performed for the $d$-wave and $p$-wave state.
Chiral superconductors, which are likely to result from either $d$-wave or $p$-wave instabilities on hexagonal lattices, are potential hosts to Majorana zero modes in their vortex cores \cite{PhysRevLett.86.268,PhysRevB.82.134521}.

{\it Experimental signatures.}
The observation of a finite $\kappa/T$ for $T\rightarrow 0$ in thermal
conductance measurements~\cite{zhao2021nodal} as well as the typical
V-shaped gap in STM measurements~\cite{Chen2021a} have provided
supporting experimental evidence for a nodal gap in AV$_3$Sb$_5$,
which would be in line with $f$-wave pairing which we obtain in a large
parameter regime for our minimal model. An $f$-wave state will have additional, clear experimental signatures.
First, since the $f$-wave state pairs electrons in the spin-triplet channel, we expect the spin susceptibility in the superconducting phase to stay constant upon lowering the temperature, which should be seen in Knight-shift measurements. A further signature of spin-triplet pairing is often a high critical field. However, recent critical field measurements for both in-plane~\cite{Du2021} and out-of-plane~\cite{zhao2021nodal} fields indicate orbital limiting at rather low fields, such that the critical fields cannot distinguish the Cooper-pair spin structure. Finally, many thermodynamic quantities allow to identify the nodal
structure through the temperature scaling of their low-temperature
behavior. With the $f$-wave order parameter being the only one with
symmetry imposed nodes, the low-energy excitations due to these nodes
can directly probe this order without requiring phase information. The
to date strongest evidence for the $f$-wave state comes indeed from
thermal conductance measurements in Ref.~\cite{zhao2021nodal}.
In order to strengthen this conclusion, other thermodynamic probes, such as the
electronic specific heat, penetration depth, or $1/T_1$ in NMR should show a square, linear, and cubic temperature dependence\cite{PhysRevB.100.014519}, respectively.
Note, however, that disorder washes out  these low-energy signatures.
Chiral $p$-wave and $d$-wave superconductivity would be in line with a possibly
highly anisotropic, but nodeless gap. Furthermore, concomitant signatures of
time-reversal symmetry breaking could be revealed through Kerr
measurements, $\mu$-SR below $T_{\rm c}$, or even new experimental
approaches such as the detection of clapping modes~\cite{poniatowski2021spectroscopic}.
Most importantly, for the specific scenario of multiple van Hove singularities, the double dome feature in the superconducting phase can tentatively be understood by the evolution of the individual van Hove bands as a function of pressure~\cite{chen2021double,Chen2021}.

Our work shows the unique principles for unconventional pairing in kagome
metals, which promises to unlock a whole new paradigm of electronically-mediated superconductivity.

{\it Acknowledgments.} This work is funded by the Deutsche
Forschungsgemeinschaft (DFG, German Research Foundation) through
Project-ID 258499086 - SFB 1170 and through the W\"urzburg-Dresden
Cluster of Excellence on Complexity and Topology in Quantum Matter - ct.qmat Project-ID 390858490 - EXC 2147. Y.I. acknowledges financial support by the Science and Engineering Research Board (SERB), Department of Science
and Technology (DST), India through the Startup Research Grant No. SRG/2019/000056 and MATRICS Grant No. MTR/2019/001042, the Abdus Salam International Centre for Theoretical Physics (ICTP) through the Simons Associateship scheme funded by the Simons Foundation, IIT Madras through the Institute of Eminence (IoE) program for establishing the QuCenDiEM group (Project$\#$ SB20210813PHMHRD002720), and the International Centre for Theoretical Sciences (ICTS), Bengaluru, India during a visit for participating in the program "Novel phases of quantum matter" (Code: ICTS/topmatter2019/12). This research was supported in part by the National Science Foundation under Grant No. NSF PHY-1748958. This project has received funding from the European Research Council (ERC) under the European Union¡¯s Horizon 2020 research and innovation programm (ERC-StG-Neupert-757867-PARATOP). The research leading to these results has received funding from the European Union¡¯s Horizon 2020 research and innovation programme under the Marie Sk\l{}odowska-Curie Grant Agreement No. 897276. The authors gratefully acknowledge the Gauss Centre for Supercomputing e.V. (www.gauss-centre.eu) for funding this project by providing computing time on the GCS Supercomputer SuperMUC at Leibniz Supercomputing Centre (www.lrz.de). Y.I. acknowledges the use of the computing resources at HPCE, IIT Madras.

\bibliography{av3sb5}

\end{document}